Graphical Abstract

**2D versus 3D-like electrical behavior of MXene thin films: insights from weak localization in the role of thickness, interflake coupling and defects.**

Sophia Tangui,Simon Hurand,Rashed Aljasmi,Ayoub Benmoumen,Marie-Laure David,Philippe Moreau,Sophie Morisset,Stéphane Célérier,Vincent Mauchamp

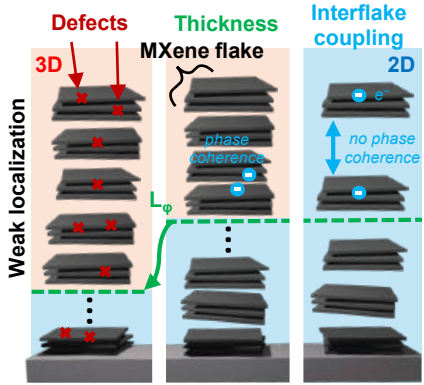

# 2D versus 3D-like electrical behavior of MXene thin films: insights from weak localization in the role of thickness, interflake coupling and defects.

Sophia **Tangui**[a], Simon **Hurand**[a,*], Rashed **Aljasmi**[a], Ayoub **Benmoumen**[a,c], Marie-Laure **David**[a], Philippe **Moreau**[c], Sophie **Morisset**[b], Stéphane **Célérier**[b] and Vincent **Mauchamp**[a]

[a]*Institut Pprime, Université de Poitiers, CNRS, ISAE-ENSMA, 11 Bd M. P. Curie, Chasseneuil-du-Poitou 86360, France*
[b]*IC2MP, Université de Poitiers, CNRS, 4 rue Michel Brunet, Poitiers 86073, France*
[c]*IMN, Nantes Université, CNRS, 2 Chem. de la Houssinière, Nantes 44300, France*

## ABSTRACT

MXenes stand out from other 2D materials because they combine very good electrical conductivity with hydrophilicity, allowing cost-effective processing as thin films. Therefore, there is a high fundamental interest in unraveling the electronic transport mechanisms at stake in multilayers of the most conducting MXene, $Ti_3C_2T_x$. Although weak localization (WL) has been proposed as the dominating low-temperature (LT) transport mechanism in $Ti_3C_2T_x$ thin films, there have been few attempts to model it quantitatively. In this paper, the role of important structural parameters – thickness, interflake coupling, defects – on the dimensionality of the LT transport mechanisms in spin-coated $Ti_3C_2T_x$ thin films is investigated through LT and magnetic field dependent resistivity measurements. A dimensional crossover from 2D to 3D WL is clearly evidenced when the film thickness exceeds the dephasing length $l_\phi$, estimated here in the $50 - 100\,nm$ range. 2D WL can be restored by weakening the coupling between adjacent flakes, the intrinsic thickness of which is lower than $l_\phi$, hence acting as parallel 2D conductors. Alternatively, $l_\phi$ can be reduced down to the $10\,nm$ range by defects. Our results clearly emphasize the ability of WL quantitative study to give deep insights in the physics of electron transport in MXene thin films.

## 1. Introduction

The possibility to go from tri-dimensional (3D) to two-dimensional (2D) materials has been one of the major breakthroughs in solid state physics and materials science, opening truly new perspective in terms of properties and applications. In this context, 2D transition metal carbides or nitrides – known as MXenes [1] – are one of the latest and largest families of 2D materials, comprising more than 40 different compounds synthesized to date, and many more predicted theoretically [2, 3]. Among all MXene compositions, $Ti_3C_2T_x$ was the first to be synthesized [4] and is still the most studied – T being the surface functionalization groups that are generated by the synthesis process, e.g., O, OH, F and Cl, and $x$ its stoichiometric coefficient, usually close to 2 although it may vary with the exfoliation condition and/or post-treatment [5] (see the structural model in Fig. 1.a). This MXene combines high metallic conductivity [6] with tunable hydrophilicity and easy dispersion in water without surfactant, enabling scalable, flexible, and cost-effective processing as thin films, inks or membranes, a major advantage for applications [7, 8]. The very good electronic conductivity of $Ti_3C_2T_x$ multilayers is key to many of the foreseen applications such as energy storage [9, 10], transparent conductive electrodes and capacitors [11, 12, 13], sensors [14] or electromagnetic interference shielding [15], among others. In order to optimize MXene multilayers electronic conductivity for target applications, a deep understanding of the physical processes governing electron transport in these complex systems is required.

Charge transport mechanisms in MXenes are governed by many different factors, one of the main being the chemistry of the transition metal sites [16]. The surface functionalization groups also impacts these properties [17, 18], by affecting the density of states at the Fermi level [19], and probably acting as scattering centers for conduction electrons thus leading to quite low elastic mean free path of typically 1 nm [20]. In addition, for a given MXene composition, electrical transport is determined by other factors like structural defects, but also the complex interplay between intra-flake and inter-flake mechanisms [21] which depends on the studied MXene and can be investigated by temperature or magnetic field dependent measurements [22, 21]. In this context, low temperature measurements have long been recognized as a highly valuable approach to investigate transport properties in diverse nanostructured materials. This is particularly true for 2D systems (e.g., thin films, 2D materials) where the quantum corrections to the Drude resistivity, which result from the interference between elastically scattered coherent conduction electronic wave functions, lead to a characteristic increase of the resistivity below some tens of Kelvins. Such phenomenon is referred to as Weak Localization (WL). Because WL is essentially an interference process, it is highly sensitive to any phenomenon altering the phase coherence between the conduction electrons and allows their detailed investigation. Such coherence loss could be due to inelastic scattering

*Corresponding author
simon.hurand@univ-poitiers.fr (S. Hurand)
ORCID(s):

a)

Ti₃C₂Tₓ

b)

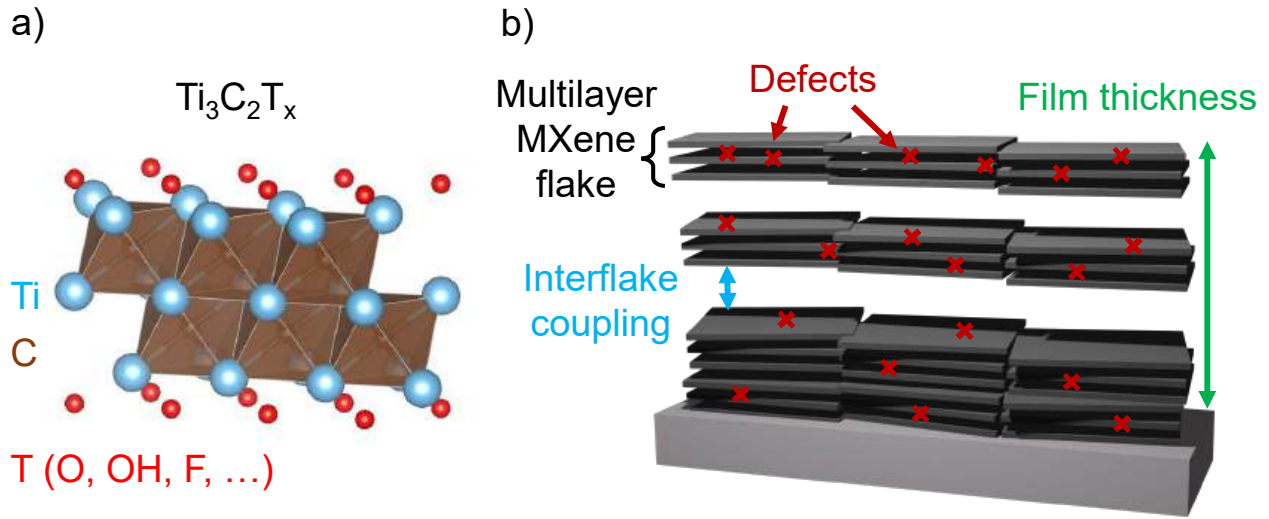

Multilayer MXene flake

Defects

Film thickness

Interflake coupling

Ti

C

T (O, OH, F, …)

**Figure 1:** a) Cristallographic structure of a $Ti_3C_2T_x$ MXene single layer with titanium in blue, carbon in brown and surface terminations in red. b) Schematic of a thin film comprising a stack of multilayer $Ti_3C_2T_x$ flakes, each of them comprising multiple $Ti_3C_2T_x$ single layers (pictured as thin black slabs). The three parameters investigated in this study that influence the dimensionality of WL are schematized, namely, the total film thickness (green), the interflake coupling (blue), and the defects (red cross).

(e.g., electron-electron or electron-phonon interactions), but also to perturbations induced by a magnetic field, either applied during the measurements or due to spin-orbit coupling (SOC) or magnetic impurities for instance [23]. In the context of 2D materials, WL has thus been used to investigate subtle effects such as SOC induced by proximity effects in graphene/WS₂ heterostructures for graphene spintronics [24], or the corrugation of graphene layers [25, 26].

In metallic MXenes, WL has been proposed as the dominating low-temperature transport mechanism. Focusing on a 28 nm-thick $Ti_3C_2T_x$ film obtained from a Physical Vapor Deposition (PVD) grown $Ti_3AlC_2$ thin film, Halim et al. [20] first reported a linear dependence of the resistance $R$ versus the logarithm of the temperature $T$ and a negative magnetoresistance (MR) below 50 K, which are two typical features of 2D weak localization. A similar behavior for $Ti_2CT_x$ etched from $Ti_2AlC$ PVD films was also reported by the same group [27]. Since then, there have been however only very few attempts to model quantitatively the resistance vs. temperature (R-T) and MR measurements. Piatti et al. [28] investigated spray-coated films and fitted the measured MR using the Hikami-Larkin-Nagaoka (HLN) model for 2D weak localization [29]. They also argued that Kondo effect or electron-electron interactions, which could also cause a linear $R$ vs. $\ln(T)$ behavior, are not consistent with the measured MR and can thus be ruled out, so that WL is indeed the dominating mechanism at low $T$. A crossover from negative to positive MR at very low temperature (< 2 K) was also observed on spray-coated $Ti_3C_2T_x$ films by Jin et al. [30], which they attributed to the appearance of spin-orbit coupling by fitting the MR with HLN 2D WL model.

However, in all these previous studies, it has been assumed a 2D character of the weak localization without further argumentation. Nevertheless, while the MXene layers are clearly 2D in structure, the transport comprises both intra- (2D) and inter- (3D) flake mechanisms. More generally, WL in layered systems shows a variety of dimensional behavior, from 2D WL to 3D isotropic or anisotropic WL that also need to be questionned in MXene thin films. In graphite for example, well-ordered graphite exhibit clear 3D anisotropic WL [31], while intercalated graphite [32] or turbostratic graphite [33] – i.e., randomly stacked graphene layers – turn to 2D WL due to the loss of out-of-plane connectivity through intercalated species or out-of-plane order, respectively. The phase coherence across neighboring layers is therefore a crucial parameter to determine whether the whole film should be treated as a single 3D anisotropic thin film, or as a stack of disconnected 2D layers acting as parallel conduction paths. This has also been investigated in topological insulator/normal insulator (TI/NI) superlattices, where the thickness of the superlattice parameter can be tuned in a highly controllable way, and thus the electronic coupling between adjacent TI layers. In Bi₂Se₃/In₂Se₃ superlattices for instance [34], the transport dimensionality switches from 3D to 2D while increasing the TI/NI thickness, demonstrating the decoupling between adjacent layers.

As a result, it is not obvious whether WL should have a 2D or 3D nature in metallic MXenes, because the flakes are separated by surface terminations and intercalated species, typically water molecules. Unraveling the dimensionality of WL in MXene thin films as a function of the main structural parameters (i.e., film thickness, interflake coupling) is thus mandatory to provide new insigths into the intra- and inter-flake transport mechanisms of this family of 2D layered materials. In addition, it is now well known that the intrinsic physical properties of MXenes are largely impacted by the level of defects in the layers which is directly related to the used exfoliation protocol. Such parameter is thus also to be investigated for a deep understanding and control of MXene thin films properties and related applications.

In this paper, we investigate the dimensionality of the low-temperature transport mechanisms in MXene. Focusing on the most conducting $Ti_3C_2T_x$ MXene, we perform low temperature and magnetic field dependent resistivity measurements on thin films of $Ti_3C_2T_x$ deposited by spin-coating, and we analyze our data in the framework of both 2D and 3D weak localization models. A schematic of the three main structural parameters influencing the dimensionality of WL in a multilayer $Ti_3C_2T_x$ thin film is shonw in Fig. 1.b. We first investigate the influence of sample thickness on the dimensionality: indeed, the 2D or 3D character is in principle determined by the comparison of the thickness $t$ of the whole film versus the dephasing length $l_\phi$ – and not versus the MXene single layer thickness which is around 1 nm [35]. We evidence a 2D to 3D dimensional crossover while increasing the thickness above typically 100 nm, indeed corresponding to the typical values of $l_\phi$ determined through the magnetoresistance measurement. As a second step, we investigate the role of interflake coupling. For this, we perform WL measurement under high or low vacuum, possibly affecting the stacking between adjacent MXene flakes, and thus modulating the interflake coupling with strong impact on the transport dimensionality. We evidence a 3D to 2D WL crossover as this coupling is weakened even for samples thicker than $l_\phi$, indicating that phase coherence between neighboring flakes is lost under low vacuum. Finally, we investigate the influence of defect engineering through ion irradiation with various fluences. We evidence a sharp reduction of the crossover thickness separating 2D and 3D WL (i.e., the dephasing length $l_\phi$) by increasing the density of defects in the films, that arises from the reduction of the conduction electrons mean free path upon irradiation.

## 2. Material and methods

### Thin film preparation

$Ti_3C_2T_x$ thin films were prepared by spin coating of an aqueous solution ($50 mg.mL^{-1}$) of delaminated MXene powder synthetized from $Ti_3C_2T_x$ powder synthetized from $Ti_3AlC_2$ precursor following the protocol detailed in [36], on top of a $1 \times 1 cm^2$ quartz substrate. Prior to deposition, the successful etching of the $Ti_3AlC_2$ MAX phase precursor in $Ti_3C_2T_x$ was checked by X-Ray Diffraction (XRD) as evidenced in the Supplementary Information (SI) Part S1. The aqueous solution was prepared through a simple hand-shaking of a water suspension of the MXene powder for 10 min, followed by sonication for 20 min. The thin films were then deposited using a Polos SPIN150i/200i spin coater at room temperature to achieve thicknesses ranging from 20 nm to 1200 nm by varying the MXene concentration in the solution.

### Thin film characterization

Thickness measurements were systematically performed on the samples by White Light Interferometry (WLI) using a Taylor Hobson Precision Talysurf CCI 6000, as shown in SI Fig. S2. Resistivity vs. temperature and magnetoresistance measurements were performed with varying temperature from 2 to 300 K and magnetic field from $-5$ to 5 T in a Quantum Design Physical Property Measurement System. Electrical contact was made by attaching gold wires to the films using silver paste in a van der Pauw configuration. Slight inhomogeneity in film thickness, as well as roughness (see SI Part S2), result in some uncertainty on the exact distribution of the current paths within the measured sample. For this reason, the absolute value of resistivity is affected by a geometrical factor and can scatter from sample to sample. However, this does not affect the relative variation of resistivity vs. temperature and magnetic field which are found to be quite robust among different samples.

## 3. Models

Weak localization has been the subject of thorough investigation for decades now. A good introduction can be found in the review of Lee and Ramakrishnan [37]. In a disordered conductor, charge carriers move freely on a length scale $l_e$ called the elastic mean free path, until they endure elastic scattering events on static defects (e.g., vacancies, impurities). When $l_e$ is way smaller than the characteristic sizes of the sample (flake width, film thickness and sample size), this is called the diffusive regime. In this case, the quantum nature of the free carriers results in destructive interferences between the electron wave functions of opposite time-reversed closed loops, leading to a backscattering contribution to the electronic transport. This implies that the electron wave function phase coherence is preserved over such loops, so that their size is limited by a dephasing length $l_\phi = \sqrt{l_e l_i}$ (also called the Thouless length), because of the inelastic scattering that destroys the phase coherence, associated to an inelastic mean free path $l_i$. The temperature dependence of the latter can be described as $l_i = aT^{-p/2}$, where $a$ is a proportionality factor, and $p$ is an exponent depending on the scattering mechanism, with $p = 1$ for electron-electron scattering and $p = 3$ for electron-phonon scattering. The integration over all time-reversed loops from $l_e$ to the cut-off length $l_\phi$ thus gives rise to a temperature-dependent negative contribution to the conductivity $\sigma$ that depends on the dimensionality of the system. In order to discuss the dimensionality of the transport mechanisms in our samples, we have used the 2D or 3D WL models describing temperature or magnetic field dependent conductivities as described below.

## Temperature dependence in 2D

For two-dimensional systems, it is the *sheet conductance* $G_s$ (in S times square, S.☐) that is affected by WL according to the following temperature dependence [37]:

$$G_s(T) = G_s(T_0) + N\,p\,G_0 \ln\left(\frac{T}{T_0}\right) \tag{1}$$

Here $G_0 = e^2/2\pi^2\hbar = 1.23 \times 10^{-5}$ S is the quantum of conductance, $e$ being the electron elementary charge and $\hbar$ the reduced Planck's constant. $G_s(T_0)$ is the residual sheet conductance in the absence of WL and $T_0$ the onset temperature of WL. Thus, $G_s$ has a linear dependence when plotted versus $\ln(T)$. In a single-layer 2D system, the slope allows an estimate of $p$. If several 2D layers are stacked vertically, but with no phase coherence of the electron wave functions between them, they simply act as parallel conducting paths and the WL contribution has to be multiplied by the number $N$ of independent layers in equation 1.

## Temperature dependence in 3D

In the case of isotropic three-dimensional systems, the *conductivity* $\sigma$ (in S.m) has a typical power-law dependence on $T$ and fitting its exponent allows a characterization of $p$ [37]:

$$\sigma(T) = \sigma(T = 0) + \frac{e^2}{\hbar\pi^3}\frac{1}{a}T^{p/2} \tag{2}$$

Regarding anisotropic 3D systems, we have to mention here that no theory is available to our knowledge for the temperature dependence, unlike the case of magnetoresistance as described below, so that equation 2 is commonly used in the literature both for isotropic or anisotropic 3D systems.

## Magnetic field dependence in 2D

For 2D systems enduring WL, in the absence of spin-orbit coupling, the variation $\Delta G_s(B) = G_s(B) - G_s(0)$ of the sheet conductance $G_s$ induced by a perpendicular magnetic field $B$ can be described by the formula from Hikami, Larkin and Nagaoka (HLN) [29, 38], which at low magnetic field can be simplified as (see SI Part S3 for the justification):

$$\frac{\Delta G_s(B)}{G_0} = N\alpha\left[\Psi\left(\frac{1}{2} + \frac{B_\phi}{B}\right) - \ln\left(\frac{B_\phi}{B}\right)\right] \tag{3}$$

Where $\Psi$ is the digamma function. $B_\phi = \hbar/4eD\tau_i = (\Phi_0/2)/\pi l_\phi^2$ is the characteristic dephasing magnetic field, corresponding to the penetration of half a quantum of magnetic flux $\Phi_0 = h/e$ within a disk of typical diameter $l_\phi$, so that the two time-reverse electron paths enclose a flux $\Phi_0$. $D = \frac{1}{2}v_F^2\tau_e$ is the diffusion constant, $v_F$ the Fermi velocity, and $\tau_e$ and $\tau_i$ the elastic and inelastic scattering times. Like for the temperature dependence, the usual HLN formula has to be multiplied by the number $N$ of phase-incoherent 2D layers acting as parallel conducting paths if the system comprises a stack of uncoupled 2D layers [39, 34]. The parameter $\alpha$ is expected to take values between 0 and 1 [29, 39].

## Magnetic field dependence in 3D

The WL correction $\Delta\sigma(B) = \sigma(B) - \sigma(0)$ to the conductivity $\sigma$ of potentially anisotropic 3D systems in a perpendicular magnetic field $B$ can be described by the theory of Kawabata [40]:

$$\Delta\sigma(B) = \alpha_{3D}G_0\sqrt{\frac{eB}{\hbar}}f_3(x) \tag{4}$$

The anisotropy parameter $\alpha_{3D} = \sqrt{D_\parallel/D_\perp}$ is the square root of the ratio between the diffusion constants $D_\parallel$ in the in-plane (within the layers) and $D_\perp$ in the out-of-plane (parallel to the magnetic field) directions. The variable $x = \hbar/4eD_\parallel\tau_i B = B_\phi/B$ is the inverse of the applied magnetic field normalized by the characteristic dephasing field for in-plane motion, and $f_3$ is the Kawabata function :

$$f_3(x) = \sum_{n=0}^{\infty} 2\left[(n+1+x)^{1/2} - (n+x)^{1/2}\right] - \left(n + \frac{1}{2} + x\right)^{-1/2} \tag{5}$$

# 4. Results and discussion

## 4.1. Transport dimensionality vs. film thickness

We first investigate the influence of sample thickness on the transport dimensionality. For this, temperature-dependent resistance (R-T) and magnetoresistance (MR) measurements were conducted on $Ti_3C_2T_x$ thin films of different thicknesses $t$. Prior to measurement, all the films were first dehydrated during 10 h under high vacuum ($P < 10^{-4}$ mbar) at room temperature inside the cryostat chamber in order to remove the water confined in the MXene thin film, thus reaching a saturating minimum value of the films' resistance.

### 4.1.1. Temperature dependent transport

Temperature dependent resistivity curves are shown in Fig. 2.a for samples with increasing thicknesses from 20 nm to 1240 nm. All the films exhibit a similar behavior: firstly, a metallic-like decrease of the resistivity with decreasing temperature is observed from 300 K until an onset temperature $T_{onset} \sim 50$ K, for which a minimum of resistance is reached. At lower temperature however, a clear resistance upturn is visible, which is a typical signature of weak localization. In order to analyze quantitatively this upturn, we first apply the 2D WL model of Eq. 1 by plotting the sheet conductance versus the logarithm of the temperature in the $2 - 30$ K range in Fig. 2.b, and a linear fitting of these plots is then performed. For the thinnest sample (20 nm), a nice linear fit is obtained as shown in Fig. 2.b. Other samples with thicknesses $t < 50$ nm have also been measured. They exhibit a similar nice linear fit, and are shown in SI Part S4.

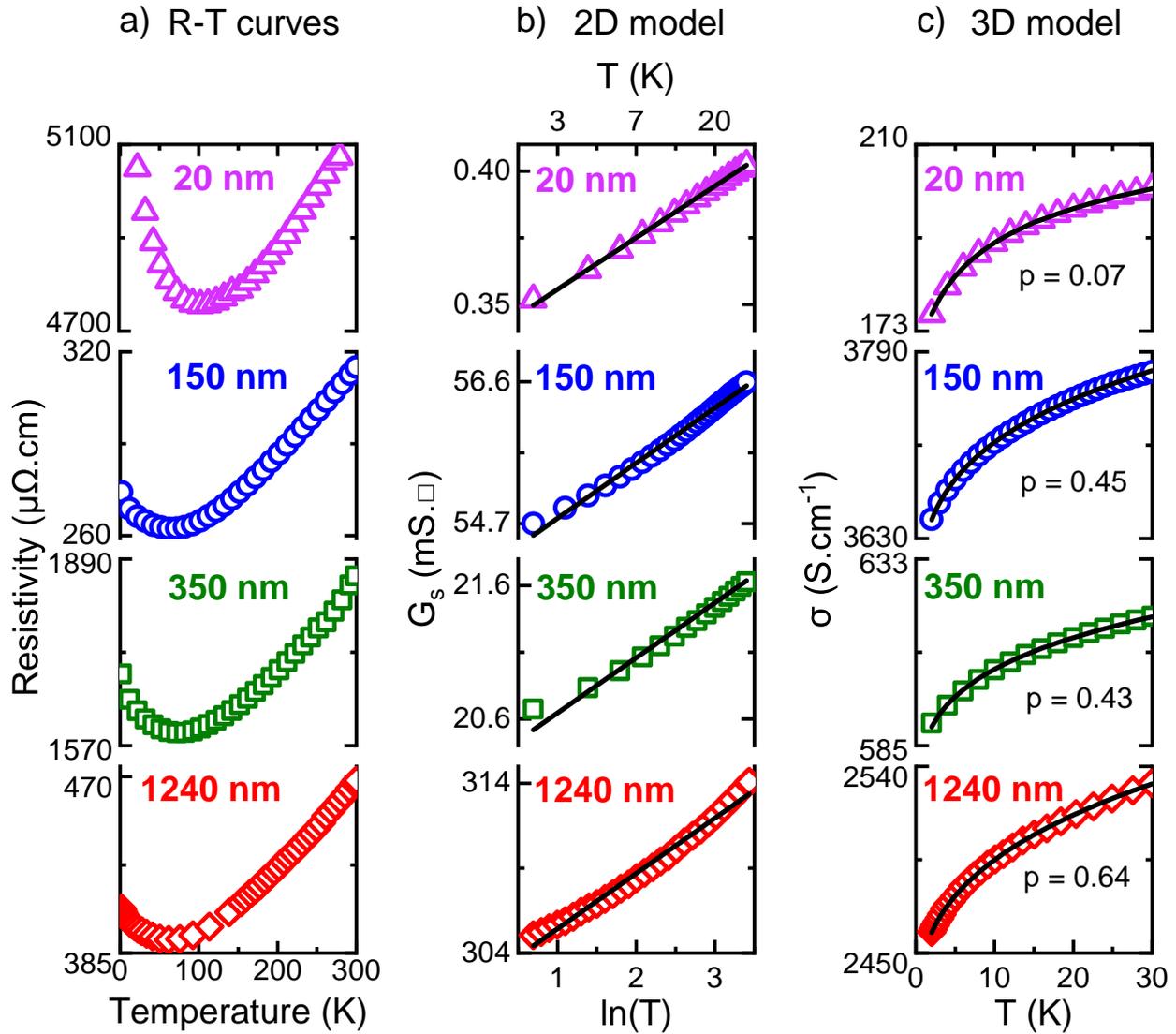

**Figure 2:** a) Temperature dependent resistivity curves for $Ti_3C_2T_x$ thin films of thickness ranging from 20 nm to 1240 nm; b) $G_s$ vs. $\ln(T)$ (2D WL); c) $T^{p/2}$ dependance of $\sigma$ (3D WL).

Therefore, the 2D WL model appears to be valid for such thin samples.

For all the thicker samples however (150 nm, 350 nm, 1240 nm), a clear departure from a linear trend is observable, so that the validity of the 2D WL model is questioned in these cases. We thus also performed a fitting of the R-T measurements using the 3D WL model of Eq. 2 instead, which predicts a power-law dependence of the conductivity versus temperature. While such a power-law is flexible enough to provide a good fit for all samples, the validity of such a fit can be checked by considering the values of the $p$ exponent hence obtained. These are reported on Fig. 2.c alongside the data and fitting curves. For all the thicker samples with $t > 100$ nm, a value of $p$ between 0.43 and 0.64 is always obtained, that is a physically meaningful value of the order of unity. On the contrary, for the thinnest samples the $p$ exponent tends to unrealistically low values (e.g. $p = 0.07$ for 20 nm). Such vanishing values are clearly not of the order of unity, and cannot have any physical meaning. To be more specific, it can be shown that the limit $p \rightarrow 0$ actually corresponds to the case $R \propto \ln(T)$, that is, the 2D WL model. Therefore, we consider the 3D WL model to be non-valid in this case. As a conclusion, the R-T measurements points out a threshold value of the thickness on the order of ~ 100 nm which determines the dimensionality of temperature dependent transport: below this value, the 2D WL model applies but the 3D WL fails; above this value, the 3D WL model is valid but the 2D WL model is not.

### 4.1.2. Magnetic field dependent transport

We further investigate the validity of 2D and 3D WL models through MR measurements on the same samples. As representative examples, the magnetic field dependence of the sheet conductance for the thinnest sample (20 nm) is displayed in Fig. 3.a, and the magnetic field dependence of the conductivity for the thickest sample (1240 nm) is displayed in Fig. 4.a. For all samples a clear positive magnetoconductance (MC) is observed, which is consistent with WL scenario. MC is always maximum at the lowest temperature, and is reduced upon increasing temperature because of the dephasing induced by the thermally driven inelastic scattering. Finally, MC disappears completely above $T_{onset} \sim 50$ K, which can be considered as a threshold temperature for the onset of WL. This is consistent with the R-T measurements, for which a resistance minimum is observed around the same temperature.

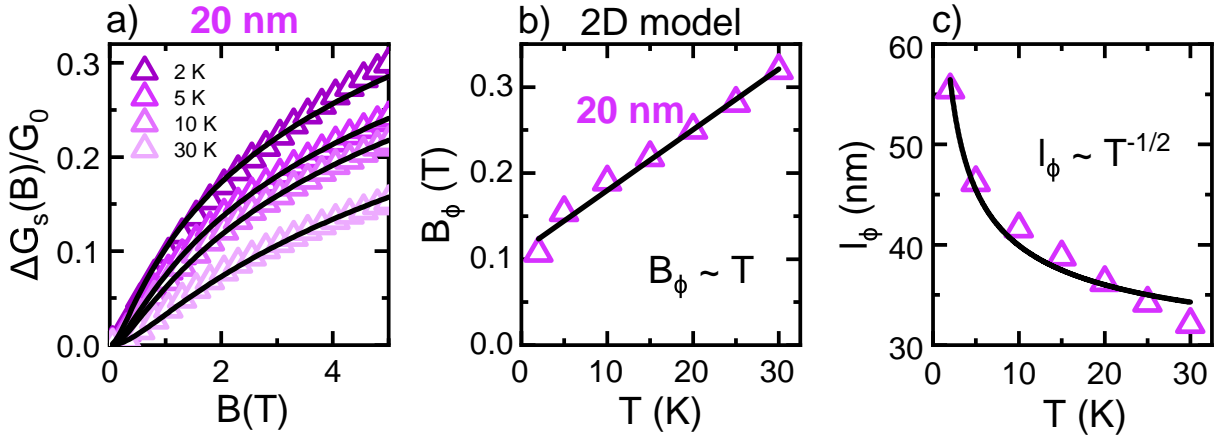

**Figure 3:** a) 20 nm-thick $Ti_3C_2T_x$ thin film magnetoconductance fitted with a 2D WL model; b) $B_\phi$ values obtained with 2D model and c) corresponding $l_\phi$ values.

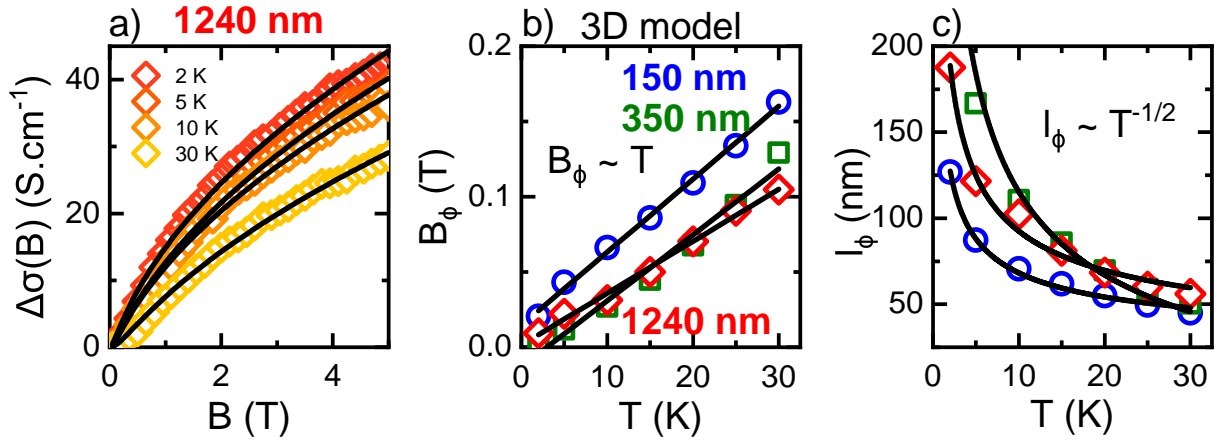

**Figure 4:** a) 1240 nm-thick $Ti_3C_2T_x$ thin film magnetoconductance fitted with a 3D WL model; b) $B_\phi$ values obtained with 3D model for 3 thick samples and c) corresponding $l_\phi$ values.

For the thinnest sample (20 nm), we apply the 2D WL model of Eq. 3 in order to fit the MC measurements, with only two free fitting parameters: the amplitude $N\alpha$ and the crossover field $B_\phi$. As can be seen in Fig. 3.a, a nice agreement is found between the model (black lines) and the experiment (symbols). The fitted values of $B_\phi$ with the 2D WL model are displayed in Fig. 3.b as a function of the temperature. $B_\phi$ follows a clear linear trend versus temperature, as evidenced by the linear fit displayed as a thick black line in Fig. 3.c. Such a linear trend indicates that inelastic scattering are dominated by electron-electron scattering with $p = 1$, so that $B_\phi \propto T$. The corresponding values of the dephasing length $l_\phi = (\Phi_0/\pi B_\phi)^{1/2}$ are reported in Fig. 3.c. The corresponding $l_\phi \propto T^{-1/2}$ tendency curve, obtained by reporting the very same parameters as the linear fit of $B_\phi$, is also displayed in Fig. 3.c, here again showing a nice agreement. $l_\phi$ always lies in a range from 40 to 55 nm: as a result, samples of thickness $t$ smaller than this range, such as the 20 nm one displayed here, are indeed expected to endure 2D WL. This is fully consistent with the R-T measurement on this sample that shows a nice linear trend of $G_s$ vs. $\ln T$ in Fig. 2.b. Therefore, confirm that samples of thickness $t < 50$ nm endure 2D WL.

Regarding the amplitude parameter, the fitted value of $N\alpha$ obtained for the 20 nm-thick sample is 0.15. For 2D system enduring WL within a single channel with $N = 1$, this leads to a value of $\alpha = 0.15$. In theory, $\alpha$ is expected to be 1 in the orthogonal case, 0 in the unitary case, and 0.5 in the symplectic case in the absence of spin-orbit coupling [29, 39]. However, these values refer to calculations performed assuming electron transport within a single band, and are thus commonly encountered experimentally in 2D semiconductors or topological insulators [41, 39] for example. On the contrary, in metals with high carrier density like $Ti_3C_2T_x$ the electron transport may involve several bands partially filled at the Fermi level, which are moreover possibly anisotropic in nature. In such cases $\alpha$ takes intermediate values between 0 and 1, that can be measured experimentally but are extremely difficult to predict theoretically. As a reference metallic material with very high carrier concentration and negligible spin-orbit contribution, one can cite the case of copper, for which an average value of $\alpha = 0.63$ has been measured experimentally [42, 43, 44] with however no theoretical confrontation.

For thicker samples, we thus apply the 3D WL model of Eq. 4 to fit the MC measurements, with here again only two free fitting parameters, namely the anisotropy parameter $\alpha_{3D}$ and the crossover field $B_\phi$. The fits show a nice agreement with the experimental data for all the thickest samples like the 1240 nm one displayed in Fig. 4.a (see SI part S5 for the 150 nm and 350 nm thick samples). It is expected that $\alpha_{3D} > 1$ because of the layered structure of the MXene film, with reduced out-of-plane diffusion constant compared to the in-plane one. However, Eq. 4 indicates that $\alpha_{3D}$ is also directly affected by any uncertainty on the absolute value of the conductivity $\sigma$. And indeed, the obtained values of $\alpha_{3D}$ (not shown) are found to lie around unity, with however some scatter resulting from some uncertainty in the precise distribution of current flow, which affects $\sigma$. As a consequence, we are not able to extract reliable information on the anisotropy from our measurements. The fitted values of $B_\phi$ are displayed in Fig. 4.b, here again showing a clear linear trend indicating electron-electron interaction inelastic scattering. The corresponding values of $l_\phi$ are displayed in Fig. 4.c, together with the $l_\phi \propto T^{-1/2}$ tendency curve. They are similar to the ones obtained by the 2D WL model displayed in Fig. 3.c, while somewhat higher, and all lie in a range from 40 to 190 nm. This implies that the thick samples with $t > 200$ nm represented here indeed verify $t > l_\phi$, and therefore are expected to endure 3D WL. Here again, this is consistent with the conclusion drawn from the R-T measurement shown in Fig. 2.c.

As a general conclusion, both R-T and MR measurements indicate a crossover from 2D WL to 3D WL for samples whose thickness $t$ is below or above the dephasing length $l_\phi$, which is shown to be of the order of $\sim$ 50 − 100 nm depending on the temperature.

## 4.2. Transport dimensionality vs. interflake coupling

The dimensionality of the electrical transport also strongly depends on the coupling between adjacent conducting channels, which can be single layer or multilayer MXene flakes. This interflake coupling can be modified by different structural parameters, like, e.g., intercalated molecules or porosity. In the present case, a simple approach based on varying the vacuum level was used to modify the level of coupling of adjacent flakes at the thin film scale.

In order to investigate this effect, R-T and MR measurements were performed on selected samples with thickness above the dephasing length determined previously, thus clearly expecting a 3D WL behavior. They were first measured under low vacuum (LV), with a $\sim$ 5 mTorr residual pressure of pure helium, right after deposition. Immediately after this, and without breaking the vacuum in the cryostat chamber, they were pumped during 10 h at room temperature under a high vacuum (HV), greater by two order of magnitude ($\sim 9{\times}10^{-2}$ mTorr), and measured under high vacuum immediately afterwards. Results for the 350 nm thick sample are discussed here, and the measurements performed on the other films are given in SI part S6.

The R-T measurement under LV shows a nice agreement with 2D WL model, visible as a clear linear trend e.g. for the 350 nm thick sample in Fig. 5.a. On the contrary, 3D WL model fit of the same measurement fails, with a very low and unphysical value of $p = 0.14$ as visible in Fig. 5.b. This is highly surprising and unexpected for such thick samples with $t > l_\phi$. One can understand this phenomenon by considering that those films behave as parallel conducting 2D layers enduring WL, with no phase coherence across them, because the adjacent MXene flakes are weakly electrically coupled in such low vacuum. The average thickness $d$ of each conducting path in the WL regime can be estimated by dividing the thickness by the linear slope of the R-T fit, leading to $d = t/Np = 350\,\text{nm}/31 = 11$ nm for example for the 350 nm sample. Given the fact that one single $Ti_3C_2T_x$ layer is around 1 nm thick, this means that each independent 2D layer – from the point of view of WL – comprises an average of 11 adjacent $Ti_3C_2T_x$ single layers. Regarding MR measurements, fitting the data on the 350 nm sample under LV with the 2D WL model gives to values of $l_\phi = 35 - 65$ nm depending on the temperature, as shown in SI Part S6. What is more, the MR fitting needs to introduce an amplitude factor $N\alpha = 9.2$, that is, far above unity. As the parameter $\alpha$ is expected to be between 0 and 1 (e.g., $\alpha = 0.15$ for the 20 nm-thick sample as shown in Section 4.1), this means that at least 9 adjacent 2D layers contribute to the electron transport as parallel conducting paths, with no phase coherence between them. The average thickness of such 2D "WL layers" can thus be estimated as $d = t/N\alpha = 350\,\text{nm}/9.2 = 38$ nm from MR measurements, which is of the same order of magnitude as the value of $d = 11$ nm estimated from the R-T data. Since both estimates of $d$ give values on the order of or smaller than the dephasing length $l_\phi = 35 - 65$ nm at low temperatures, this provides a reasonable explanation for the observed 2D WL behavior, even if the sample thickness $t$ is greater than $l_\phi$.

Thereafter performing R-T measurement upon HV, the experimental data noticeably deviate from the linear behavior of the corresponding 2D WL model, as can be seen in Fig. 5.c. Very consistently, a 3D WL model gives $p = 0.6$, that is a much more physically meaningful value of the order of unity, as visible in Fig. 5.d. Thus, applying HV has the effect to improve the interflake coupling in our MXene thin films, possibly because of an improved stacking between adjacent flakes. Therefore, applying HV apparently restores the phase coherence across the adjacent layers up to a thickness equal to the dephasing length $l_\phi$, so that thick samples endure 3D WL as observed in the previous Section 4.1, in which all the measurements have been performed under HV.

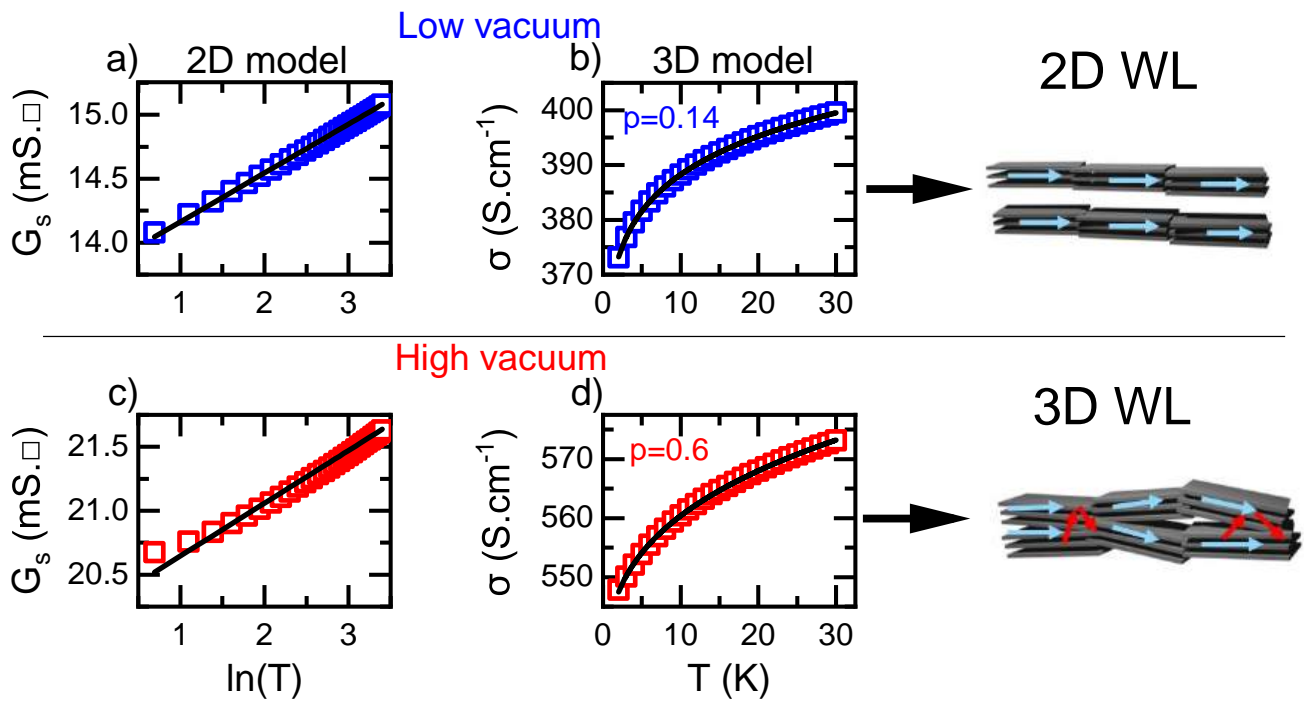

**Figure 5:** Low vacuum vs. high vacuum resistivity measurements on a 350 nm-thick $Ti_3C_2T_x$ thin film. In each case, the 2D model a) and c) ($G_s$ vs. $\ln(T)$) and 3D model b) and d) ($T^{p/2}$ dependence of $\sigma$) of WL are explored. The schematics on the right illustrate the two possible scenarios: (top) uncoupled MXene flakes, where electron phase coherence is preserved only within individual multilayer flakes, resulting in 2D WL; (bottom) coupled MXene flakes, where electron phase coherence is restored between neighboring flakes, resulting in 3D WL. Intra- and inter-flake electron motion preserving phase coherence is pictured as blue and red arrows, respectively.

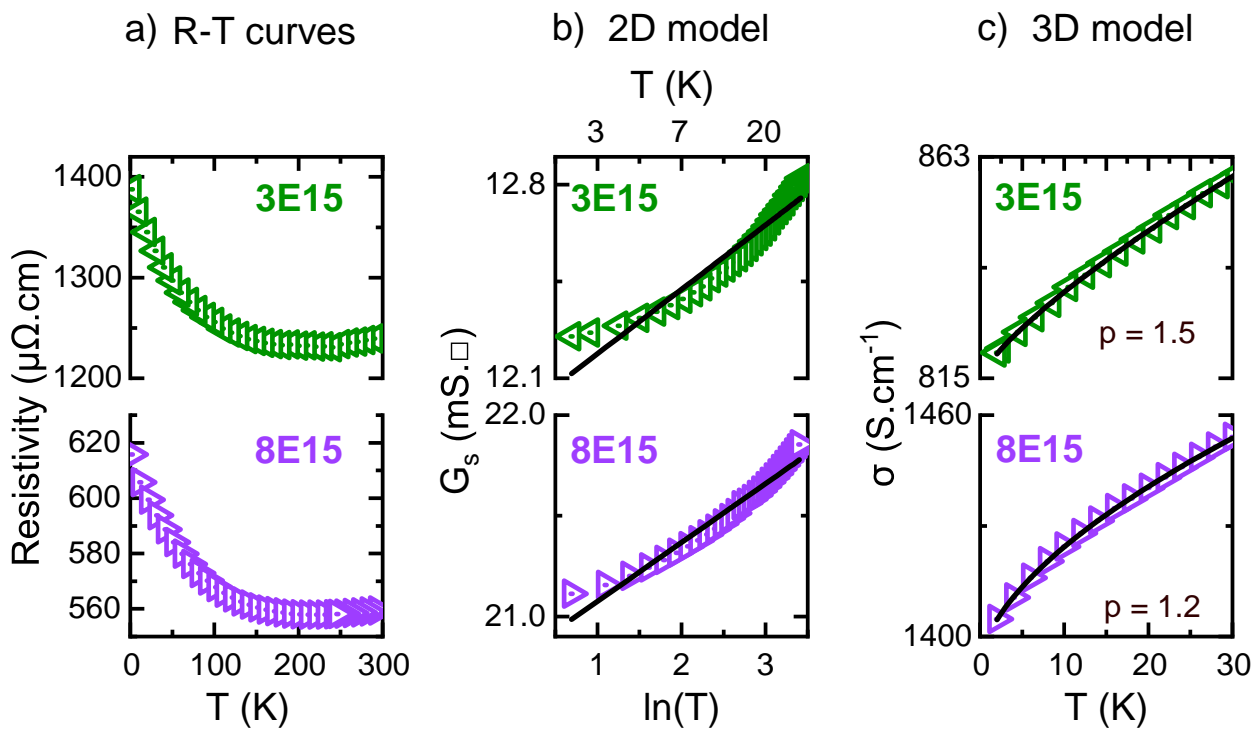

**Figure 6:** a) Temperature dependent resistivity curves for $Ti_3C_2T_x$ irradiated thin films 3E15 and 8E15; b) $G_s$ vs. $\ln(T)$ (2D WL); c) $T^{p/2}$ dependence of $\sigma$ (3D WL).

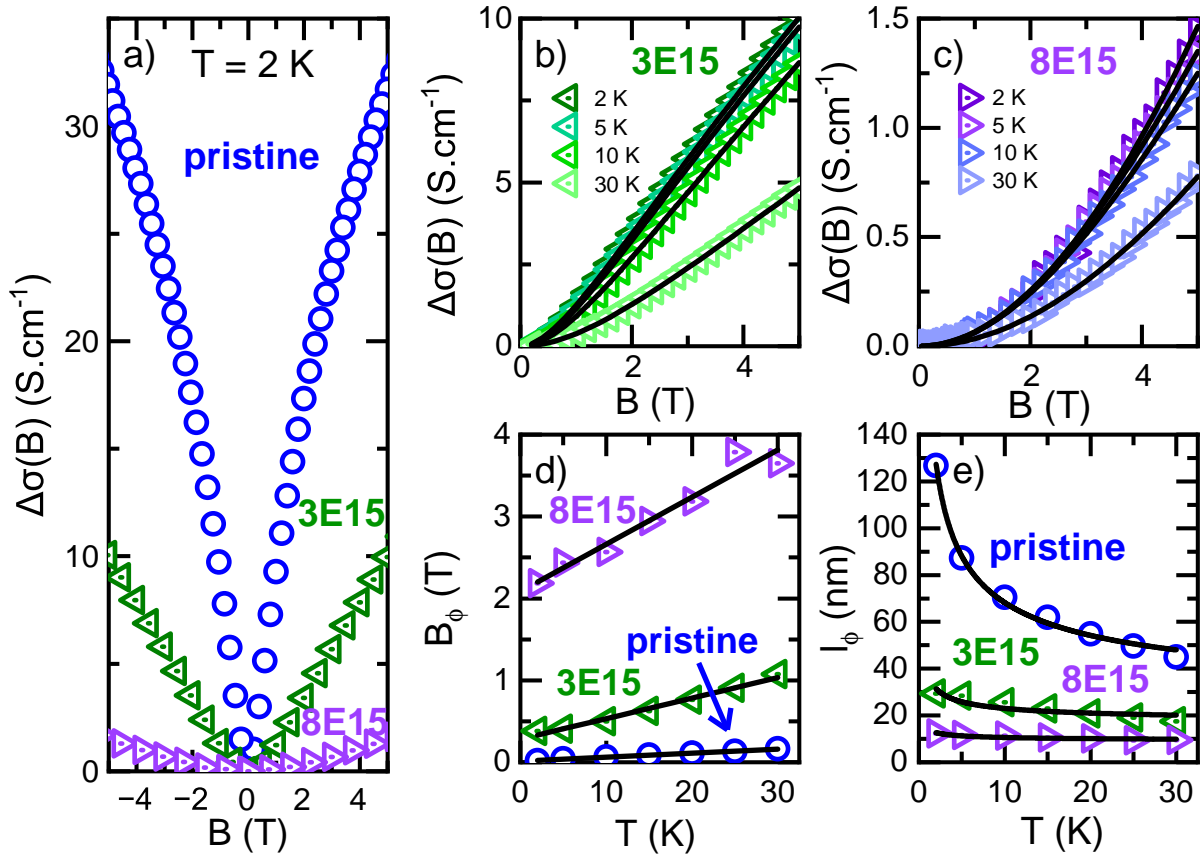

**Figure 7:** a) Magnetic field dependent conductivity curves for $Ti_3C_2T_x$ 150 nm-thick pristine film and irradiated films 3E15 and 8E15; b) 3E15 and c) 8E15 irradiated film magnetoconductance fitted with a 3D WL model; d) $B_\phi$ values obtained with 3D model for the pristine and 2 irradiated samples and e) corresponding $l_\phi$ values.

As a conclusion, weak interflake coupling results in 2D WL: samples with thickness above $l_\phi$ are then split into decoupled 2D "WL layers", thinner than $l_\phi$, comprising some tens of MXene sheets. The phase coherence of the electron wavefunction is maintained within each layer, but not across them. There is therefore a dimensional crossover from 2D to 3D WL upon weakening the interflake coupling. These results highlight the prominent role of interflake coupling in the physics of electron transport in MXene thin films.

### 4.3. Transpot dimensionality vs. defects

Having unravelled the effect of thickness and interflake coupling on the dimensionality of the electronic transport, we further investigate the influence of the defects on the 2D to 3D WL crossover. This can be achieved in a controllable and tunable way by using ion irradiation, as proposed by Benmoumen et al. [45]. In this work, ion irradiation was performed on $Ti_3C_2T_x$ spin-coated thin films identical to those investigated here, with Helium ions accelerated at an energy of 180 keV. Doing so, structural damage is localized in the first $\sim 200$ nm of the film, with negligible implantation as the helium ions end up far within the quartz substrate. Irradiation conditions were chosen in order to avoid any doping or chemical effect, and the level of damage is determined by the ion fluence used to irradiate the sample. The main results of this work are that increasing the ion fluence allows increasing the amont of structural disorder within the $Ti_3C_2T_x$ layers. In addition, one can remove the intercalated water layers (low fluence), modify the surface functionalization (low to medium fluence) and even induce Ti sputtering (high fluence). We here investigate two of the films analyzed in Ref. [45], namely one sample irradiated at $3 \times 10^{15}$ ions.cm$^{-2}$ which is 70 nm thick (hereafter named 3E15), and a second one at $8 \times 10^{15}$ ions.cm$^{-2}$ which is 150 nm thick (hereafter named 8E15). For the structural and chemical characterizations of these films, the reader is refered to Ref. [45]: they correspond to the two lowest fluences, well below the Ti sputtering threshold so that the $Ti_3C_2T_x$ MXene layers composition is preserved but with a reduction of the number of intercalated water molecules, an increase of the O content in the T-groups and an increase of the structural disorder within the $Ti_3C_2T_x$ skeleton.

R-T measurements performed under high vacuum of samples 3E15 and 8E15 are shown in Fig. 6.a. As can be seen, a clear increase in $T_{min}$, for which the minimum of resistance is reached, is noticeable upon irradiation, with $T_{min} = 210$ K for 3E15 and $T_{min} = 200$ K for 8E15, as compared e.g. to $T_{min} = 50$ K for the pristine sample of thickness 150 nm displayed in Fig. 2.a. This is accompanied by a sharper increase of resistance at lower temperatures: the resistivity residual ratio RRR, defined as RRR $= R(2 \text{ K})/R(300 \text{ K})$, increases strongly, from 0.88 for the 150 nm-thick pristine sample to 1.11 for 3E15 and 1.10 for 8E15, that is becoming greater than 1 for irradiated samples. All these observations can be explained by an increase of the residual resistivity at low temperature, meaning that irradiation indeed results in an increase of the concentration of static defects within the MXene films.

In the case of such disordered systems, a change in the transport mechanism may be suspected, for instance, to thermal activation (TA) or variable-range hopping (VRH) behavior, as was observed by Halim et al. [22] in Mo-based MXene. However, analysis of the R-T measurements within TA or VRH models were proved to fail, as shown in SI Part S7, so that we expect WL to remain the dominating mechanism inducing the insulating-like upturn of resistance. We therefore apply 2D and 3D WL models to the R-T measurements, as shown in Fig. 6.b and c. Remarkably, 2D WL seems to fail drastically for 3E15 and 8E15 with a clear departure from linear behavior in Fig. 6.b. This is somewhat surprising, especially for 3E15, since this sample has a rather low thickness of 70 nm, of the same order as the previously estimated values of $l_\phi$, so that it could be expected to endure 2D WL. On the contrary, 3D WL model provides a nice fit to these samples as can be seen in Fig. 6.c, with a rather physically meaningful fitted value of $p = 1.5$ (3E15) and 1.2 (8E15).

MR measurements were then performed on these samples, and displayed in Fig. 7. Strikingly, we observe a sharp decrease of the overall magnetoconductance amplitude upon increasing the fluence, as can be seen in Fig. 7.a. For 8E15, the typical S-shape of WL MR curves even eventually collapses into a simple parabolic behavior. Analysis of these MR measurements in the framework of 3D WL model are displayed in Fig. 7.b and c. The fitted values of $B_\phi$ and correspondingly $l_\phi$ are plotted in Fig. 7.d and e, together with the fitted values for the 150 nm-thick pristine sample. A sharp increase of $B_\phi$ is observed upon increasing fluence, meaning that the inflection point of the S-shape MR curve is shifted towards higher $B$ value, until it is being pushed close to the limit of the measurement range, namely 5 T, for sample 8E15, resulting in the simpler parabolic MR curves of Fig. 7.c. For both 3E15 and 8E15 samples, $B_\phi$ retains its linear dependence on temperature, indicating electron-electron interaction limited inelastic scattering, like in the case of pristine samples discussed in Section 4.1. Regarding $l_\phi$, a sharp decrease is observed upon irradiation, from 70 nm for the pristine sample down to 30 nm for 3E15 and 15 nm for 8E15 for example at $T = 10$ K with the 3D

WL model (Fig. 7.d). This clearly illustrates that structural defects, as induced by ion irradiation in the present case, tend to lower the coherence length $l_\phi$ so that the corresponding MXene thin films turn to exhibit a 3D like behavior at much lower thicknesses – i.e., twice lower for the 3E15 sample and four times lower for the 8E15 one – as compared to the pristine thin films. Structural defects, as usually observed in layers etched in harsh conditions [46], have thus a critical impact on electron transport possibly leading to a 3D-like behavior event for very thin MXene films.

The origin of such a dramatic effect of irradiation on $l_\phi$ can be understood as follows. Ion irradiation leads to an increase of local defects in the film, and thus to a reduction of the elastic mean free path $l_e$. And indeed, a decrease of 60 % and 65 % of the electron mobility $\mu \propto l_e$ has been observed by Benmoumen et al. [45] in these 3E15 and 8E15 samples respectively. As $l_\phi = \sqrt{l_e l_i}$, this would lead to a decrease of 40 % of $l_\phi$. This is a sharp reduction, although not quite reaching the even sharper reduction of 80 % observed between the pristine and 8E15 samples. A similar decrease of $l_\phi$ upon introduction of chromium impurities in an oxide heterointerface has been observed by Singh et al. [47] and also attributed to the decrease of $l_e$.

## 5. Conclusions

We have questioned the claim of 2D WL explanation for the upturn of resistance in $Ti_3C_2T_x$ MXene thin films, and evidenced a crossover to 3D WL when the thickness of the films becomes higher than the dephasing length $l_\phi$ that we have estimated in the $50-100$ nm, depending on the temperature. This crossover can further be controlled through the interflake coupling, here controlled by the level of vacuum impacting intercalated species and porosity: weakening the coupling between adjacent flakes results in the loss of the electron wave function phase coherence across them. Each flake thus acts as a parallel conducting path enduring 2D WL: its thickness has been estimated here to lie in the 10 nm range, way smaller than $l_\phi$, so that 2D WL can even be recovered for samples thicker than $l_\phi$. On the other hand, we have shown that defects dramatically reduce $l_\phi$, down to 10 nm at the highest irradiation dose of $8 \times 10^{15}$ ions.cm$^{-2}$ studied here, and imposes 3D WL down to very low thicknesses, namely few tens of nanometers.

We emphasize that the models used here to study the transition from 2D to 3D WL are not restricted to the $Ti_3C_2T_x$ MXene, and may be applied more generally to any material comprising a stack of conducting 2D layers enduring WL. This includes other metallic MXene compositions (e.g., $Ti_2CT_x$, $Nb_2CT_x$, among others), but also other materials such as topological insulators heterostructures [34], etc. In particular, the transition from 3D to 2D WL upon increasing the out-of-plane disorder or interlayer spacing in graphite has been evidenced by Piraux et al. [32] and Bayot et al. [33]. On the contrary, these WL models do not apply to some MXene compositions where the transport properties are rather of semiconducting or insulating nature:

for example, Halim et al. [22] evidenced the relevance of variable range hopping (VRH) models to study the low-temperature transport properties of Mo-based MXenes.

These results demonstrate the potential of studying WL in MXene thin films for a fine understanding of their transport properties. The impact of intercalated species on the WL could be probed thanks to WL measurements and models to unravel the dimensionality of the electron transport in custom-made intercalated films, which are envisioned in a wide range of application such as electromagnetic interference shielding for instance. Also, doping or decorating MXenes flakes with magnetic impurities or atoms with strong spin-orbit coupling (e.g., Au or Pt for catalysis applications) can possibly induce spin-orbit coupling as an additional electron scattering phenomenon, for which WL is known to be a highly sensitive probe [23]. Overall, these results pave the way to a finer characterization and control of the transport properties of MXenes thin films which are determined by both intra- (2D) and inter- (3D) flake electron transport, and demonstrate the potential of the study of WL for achieving such a goal.

2D versus 3D-like electrical behavior of MXene thin films: insights from weak localization in the role of thickness, interflake coupling and defects.